\documentclass[slac_one]{revtex4}
\usepackage{graphicx}
\usepackage{fancyhdr}
\begin{document}

\title{Prospects for standard SUSY searches at the LHC} 

%

\author{Sascha Caron (for the ATLAS and CMS collaborations)}
\affiliation{Physikalisches Institut, Universit\"at Freiburg, Germany}

\begin{abstract}
We present the 
recent prospects and strategies of the ATLAS and CMS experiments
in the search for signatures of Supersymmetry. The emphasis is placed on
standard signatures with missing transverse energy and on the initial data
taking periods.  
\end{abstract}

\maketitle

\thispagestyle{fancy}

\section{INTRODUCTION} 
The Large Hadron Collider with its ATLAS and CMS detectors will soon provide
us with data to explore a 
new energy scale. 
Although already the very first data may contain surprises the first concluding results for the
search for new physics are expected with the full 2009 data, where the LHC
is expected to operate at $\sqrt{s}=14$~TeV. 
This article summarises the recent prospects for Supersymmetry (SUSY)
with this data for the CMS and ATLAS experiments.  
In the majority of cases an integrated luminosity 
of $1 \rm{fb}^{-1}$ is assumed.

Supersymmetry is one of the most favoured candidates for new physics. 
In this article we concentrate on the ``standard'' assumptions for SUSY,
i.e. we assume that the R-parity quantum number is conserved. 
If this assumption is true, new SUSY particles would be produced in pairs and 
each decays into states which include the lightest SUSY particle (LSP). 
Other SUSY models (``non-standard'') are discussed elsewhere in 
these proceedings \cite{nonstandard}. 
The LSP is only weakly interacting due to cosmological arguments and
leads to the most characteristic feature of these SUSY events, missing transverse momentum.
A general search strategy for standard SUSY signatures would be
the selection of events with large missing transverse energy and reconstructed 
particles with large transverse momentum. 
At the LHC these objects are predominantly 
jets since 
the coupling strength of the strong force would cause an abundance of squarks and
gluinos if these particles are not too heavy.
Squarks or gluinos will cascade 
decay to jets, some number of leptons or photons depending on the SUSY parameters and missing transverse momentum 
caused by the LSPs.
The searches for generic SUSY signatures with R-parity conservation are
performed by searching for more events than expected in a countable 
number of different channels. These channels explore a large variety of possible signals, e.g. 
ATLAS studied various different jet (2,3,4) and lepton (0,1,2,3) multiplicities, channels with taus and photons 
~\cite{atlas}. CMS studied channels with jets and 0,1,2 leptons, 3 leptons, $Z$, top, photons or a Higgs boson\cite{cms}.
The main challenge in these searches is to 
reliably control the Standard Model background expectations for our new experiments.
In the following we assume mostly
 the 5-parameter mSUGRA as a ``general'' model
for R-parity conserving SUSY. Both ATLAS and CMS have defined a set
of benchmark points. We will show in the following examples for CMS point LM1 and
for ATLAS point SU3.
CMS point LM1 ($m_0=60$~GeV, $m_{1/2}=250$~GeV, $\tan \beta =10$, $A_0=0$~GeV, $\mu=+$) has a cross section at 14 TeV of
55 pb and squark (gluino) masses of around 560 (610) GeV.
ATLAS point SU3 ($m_0=100$~GeV, $m_{1/2}=300$~GeV, $\tan \beta =6$, $A_0=-300$~GeV, $\mu=+$) has a cross section at 14 TeV of
28 pb and squark (gluino) masses of around 630 (720) GeV.
We will first present a 'discovery strategy' based on inclusive
searches for signals with missing transverse momentum and finally show
examples of  measurements of observables sensitive to model parameters.

\section{INCLUSIVE SEARCHES FOR SUSY SIGNALS}
\begin{figure*}[t]
\centering
\includegraphics[width=50mm]{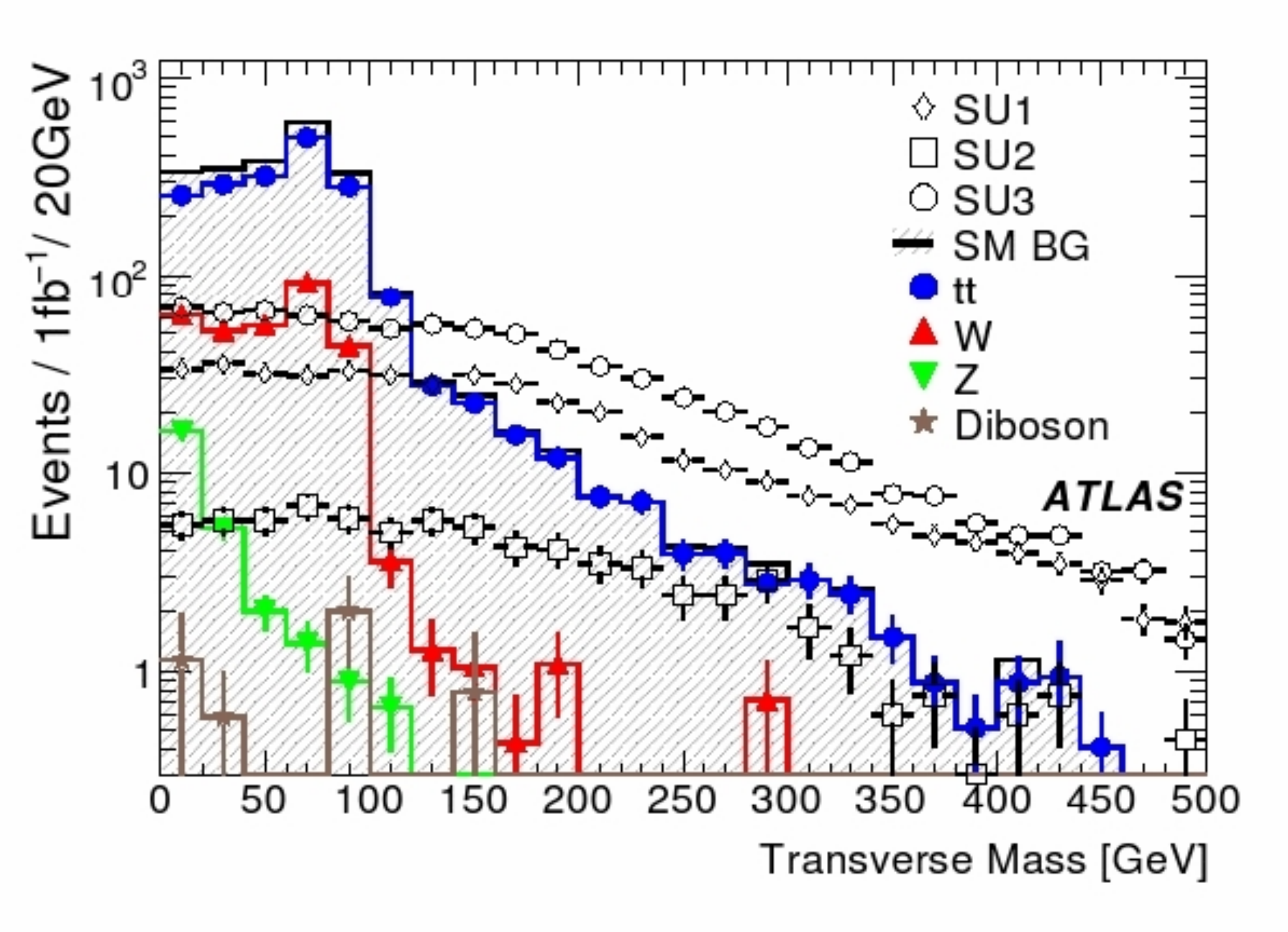}
\includegraphics[width=50mm]{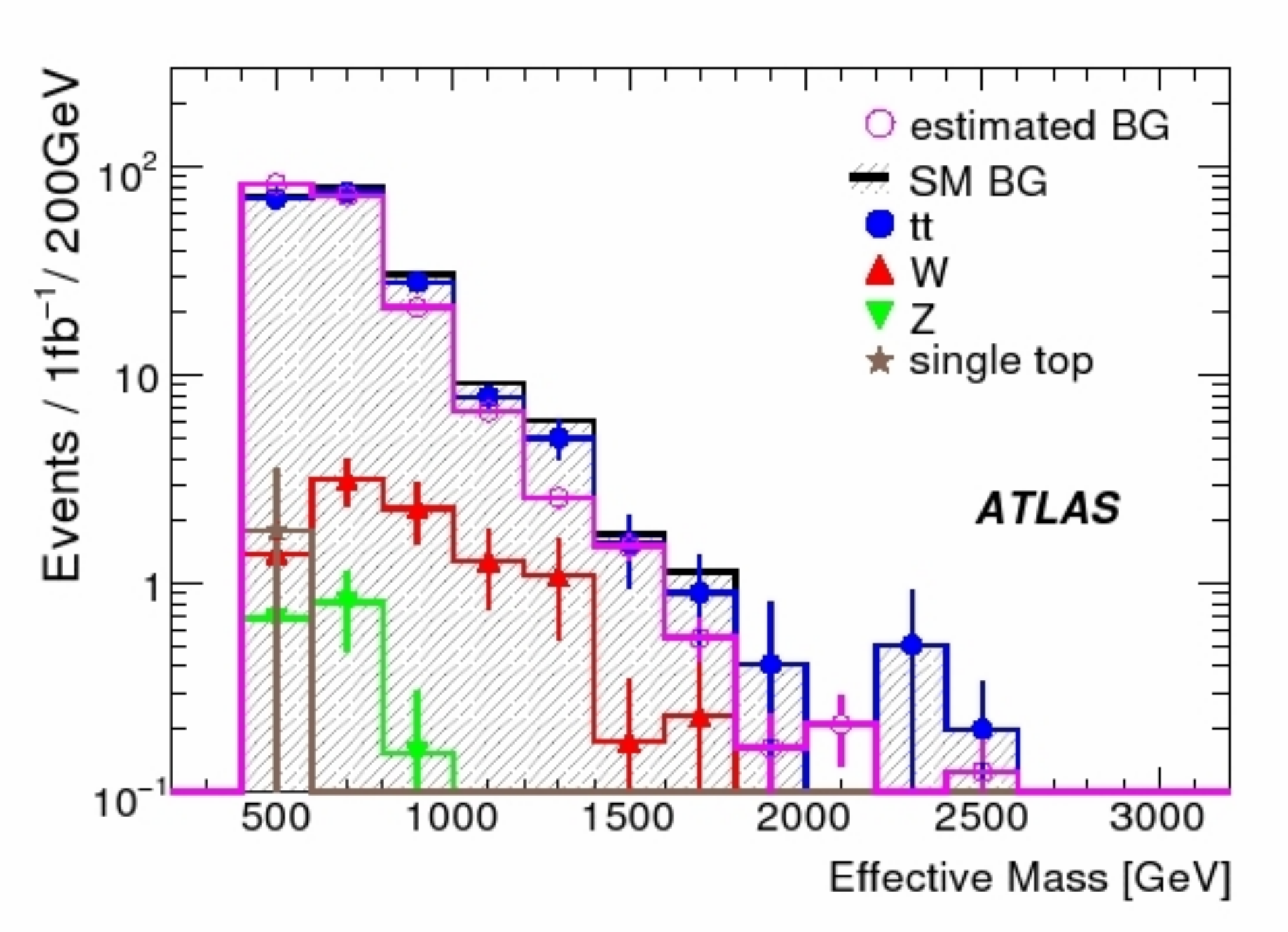}
\put(-245,75){preliminary}
\put(-93,75){preliminary}
\caption{The transverse mass $M_T$ distribution (left) and the estimated  $M_{\rm{eff}}$ distribution of $t\overline{t}$ events for the ATLAS 4-jet 1-lepton channel (right).} \label{mt}
\end{figure*}

\begin{figure*}[t]
\centering
\includegraphics[width=55mm]{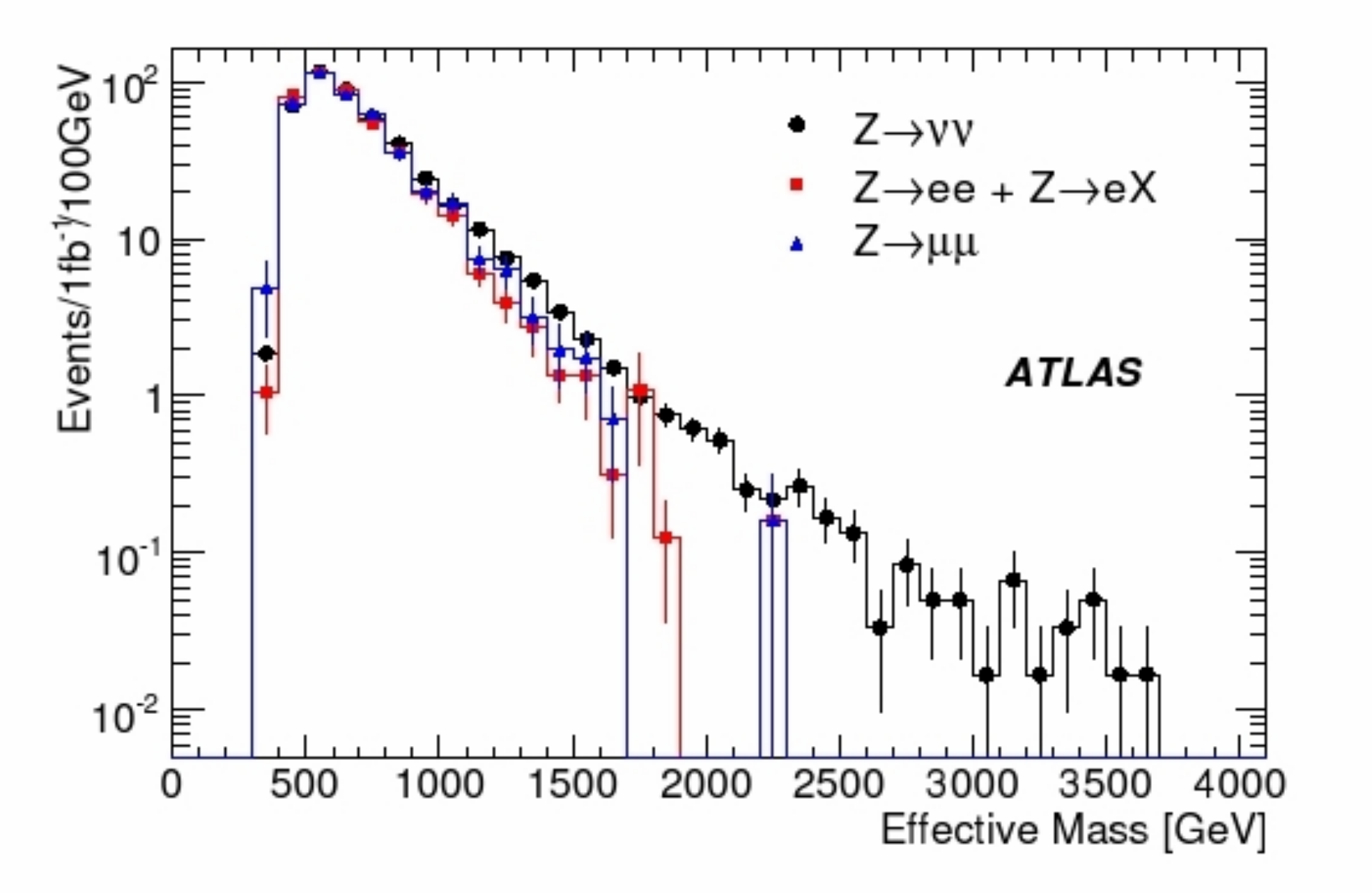}
\includegraphics[width=45mm,height=40mm]{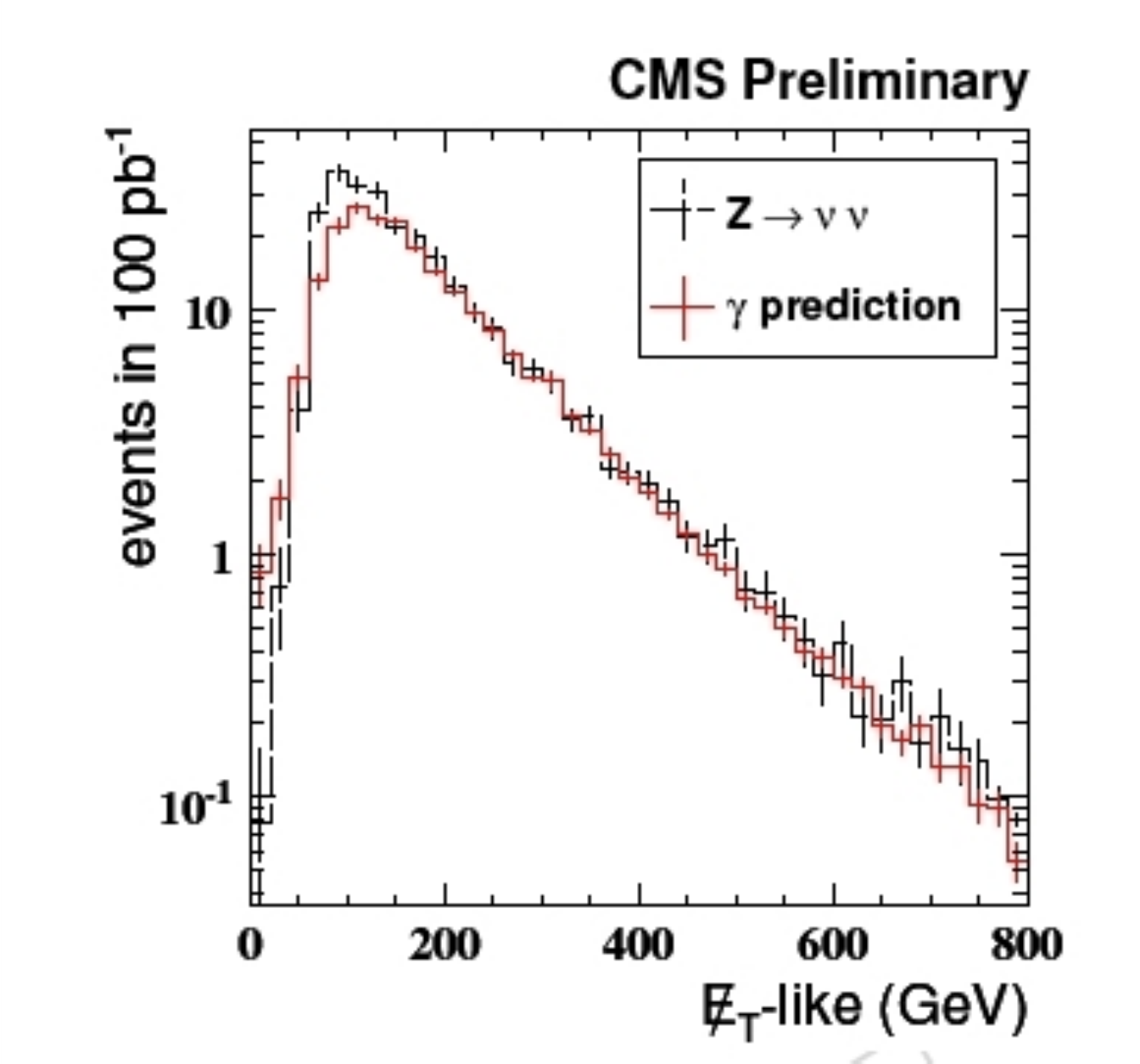}
\put(-190,44){preliminary}
\caption{The $M_{\rm{eff}}$ distribution for $Z\rightarrow \nu \overline{\nu}$ events and estimated from $Z \rightarrow e \overline{e} $ and $Z \rightarrow \mu \overline{\mu} $ events at ATLAS (left). The $E_{T,\rm{miss}}$ distribution 
from  $Z\rightarrow \nu \overline{\nu}$ and estimated from photon+jets events at CMS (right).} \label{z}
\end{figure*}

\begin{figure*}[t]
\centering
\includegraphics[width=57mm]{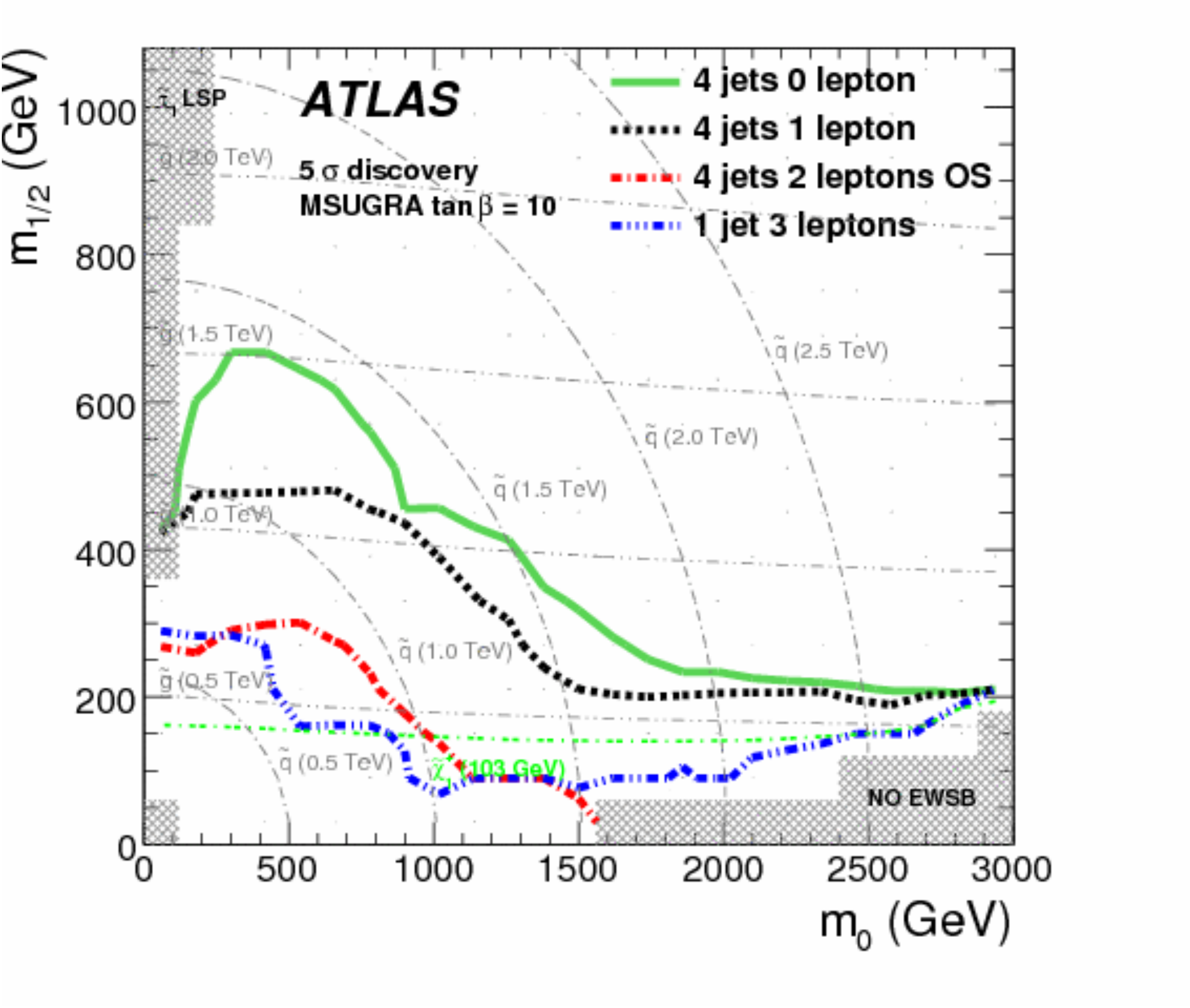}
\includegraphics[width=67mm]{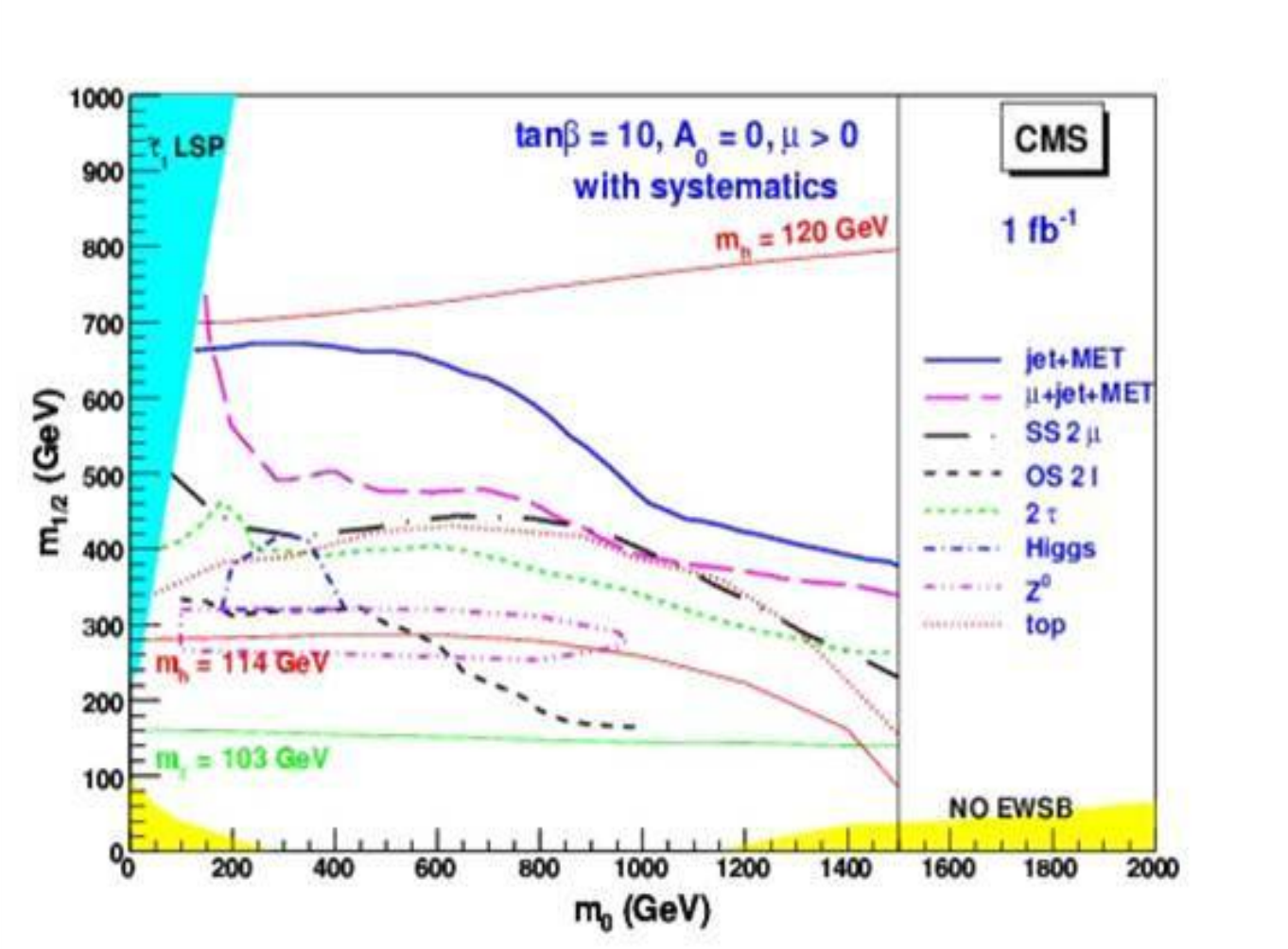}
\put(-285,75){preliminary}
\caption{The 1 $\rm{fb}^{-1}$ 5$\sigma$ mSUGRA contours for the ATLAS and CMS experiment as a function of $m_0$ and $m_{1/2}$.} \label{reach}
\end{figure*}

The basic idea in an inclusive search for SUSY signals 
is to discover a significant deviation between data and expectations  
in a dataset where signals of SUSY events are expected (signal regions).
\begin{figure*}[t]
\centering
\includegraphics[width=70mm]{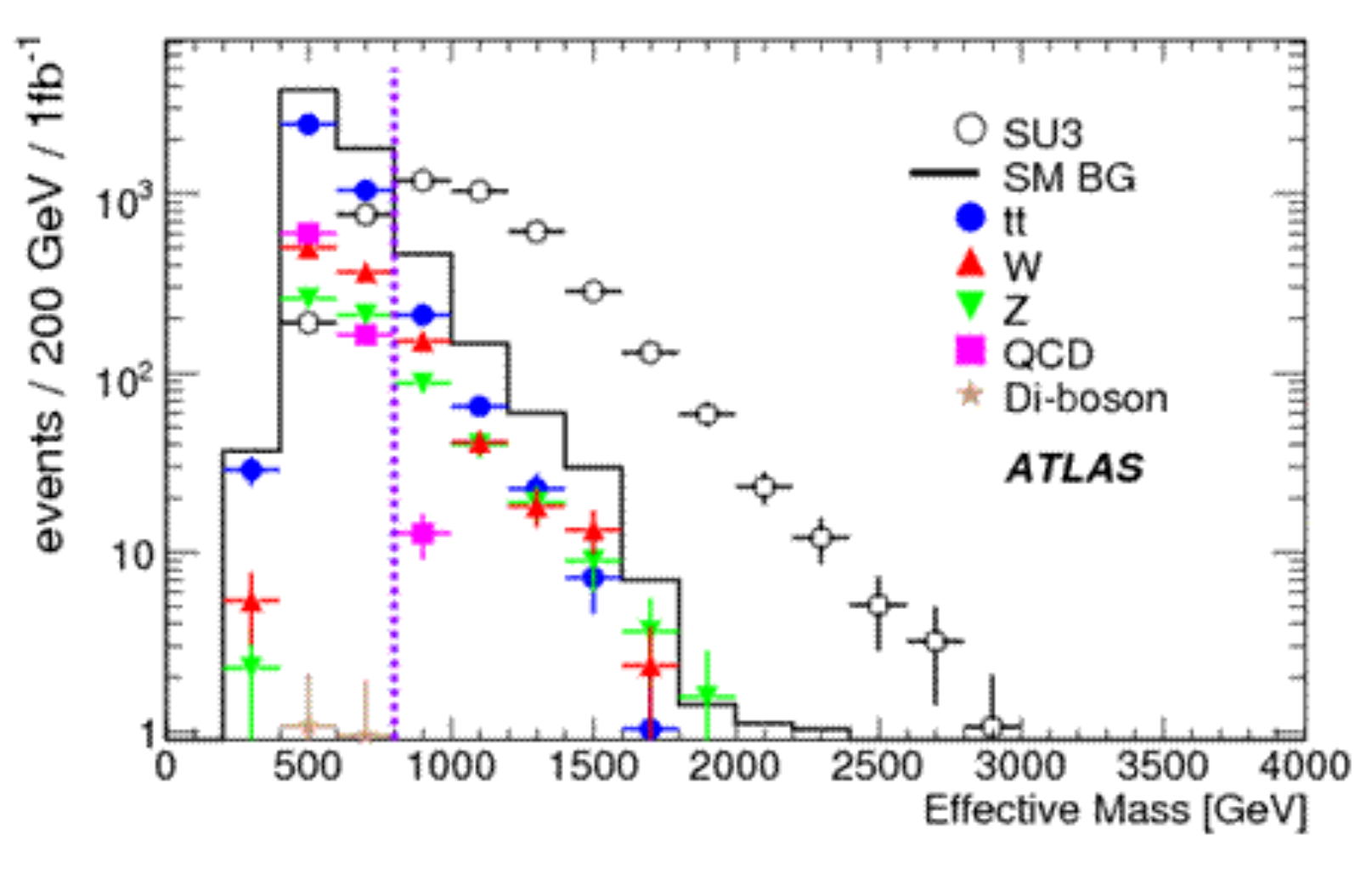}
\includegraphics[width=50mm,height=50mm]{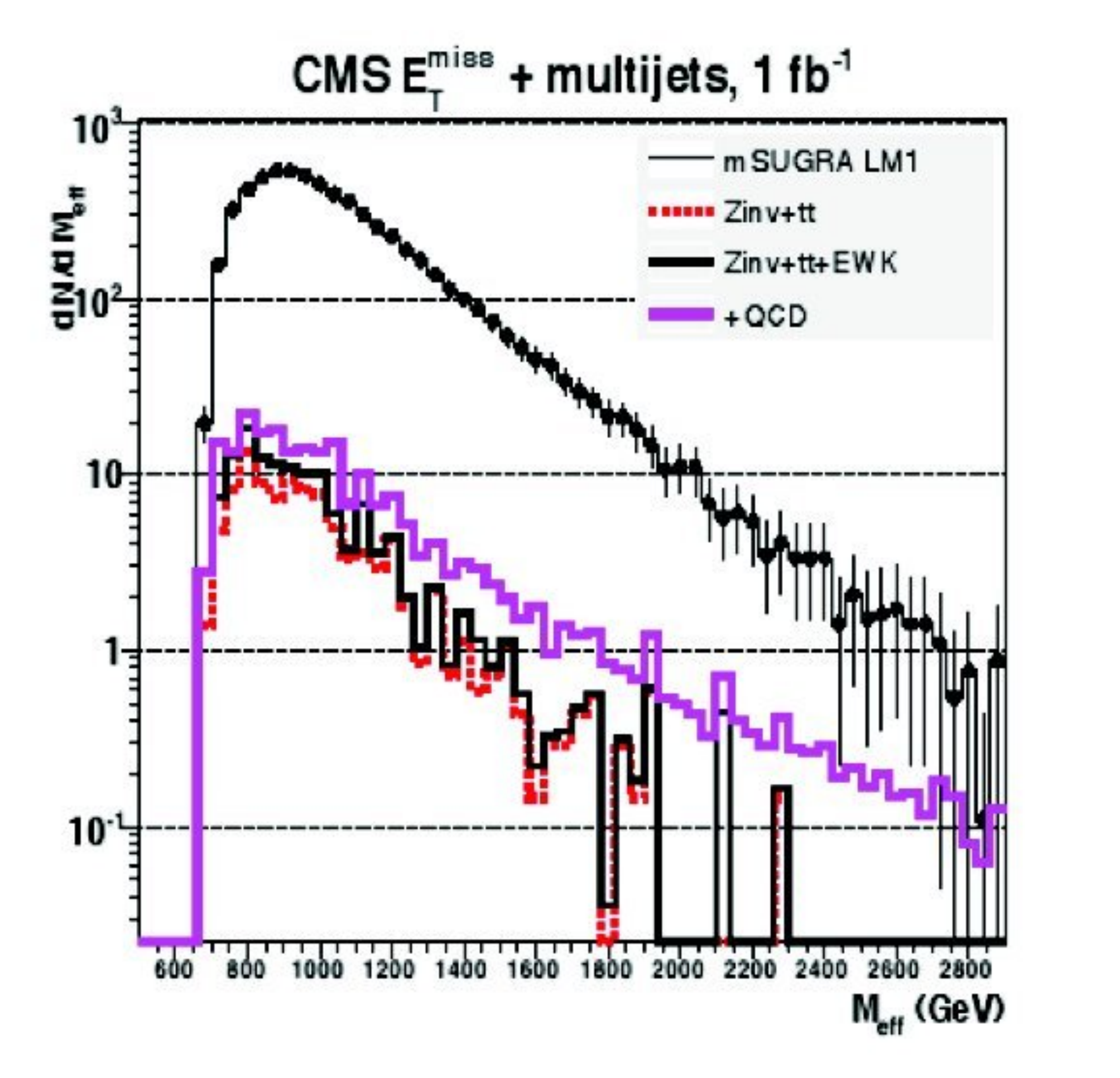}
\put(-205,50){preliminary}
\caption{The effective mass $M_{\rm{eff}}$ distribution after final selection for the inclusive jet channel at ATLAS (left) and CMS (right).} \label{meff}
\end{figure*}
In the following we briefly describe some selections of the ATLAS 
and CMS experiments and discuss then their discovery reach in SUSY parameter space.  
The data selection has been optimised to reduce the number of events from Standard Model processes using
Monte Carlo sets fully simulated with the detector response based on Geant 4 for both signal and background processes.
Figure~\ref{meff} shows the effective mass $M_{\rm{eff}}$ distribution of the ATLAS 4-jet 0-lepton selection
and the baseline CMS 3-jet selection. $M_{\rm{eff}}$ is here 
defined as the sum of the $p_T$ of the jets
and the missing transverse momentum.
ATLAS requires at least four jets with $p_T>50$~GeV with at least one with $p_T>100$~GeV. CMS requires
the 3 jets with $p_T$ cuts of $180$, $110$ and $30$~GeV. 
In both experiments the jet channels are made orthogonal to the lepton selections by vetoing events with
isolated leptons (and isolated tracks at CMS).  
The missing
transverse momentum was required not to point in the direction of jets in $\phi$ and to exceed 
$0.2 M_{\rm{eff}}$ and $100$~GeV at ATLAS and 200 GeV at CMS.  
As can be seen in Figure~\ref{meff} the background from QCD events can 
be reduced to a fairly small number.
The main background comes from top pair and $W$ production where the lepton was not reconstructed and 
the irreducible background of $Z\rightarrow \nu \overline{\nu}$ events.
Both Monte Carlo predictions and the detector response have largely unknown uncertainties for new
experiments at LHC energies. These analyses are also exploring shapes, tails of energy distributions and
high jet multiplicities. All this makes the use of control measurements for these backgrounds a
basic necessity. Both ATLAS and CMS have developed various methods to estimate and control the background expectation
with data. In the following a few examples are discussed.
\subparagraph*{Background from QCD events}
Large missing transverse energy in QCD events originates mostly from jet mismeasurements or calorimeter effects.
If one jet is mismeasured the $\phi$ coordinate of the missing transverse momentum is close
to $\phi$ of the jet. A cut in the $| \phi_{\rm{jet}} - \phi_{\rm{miss}} |$ plane is found to be useful in removing QCD events
by demanding high values and it can provide a QCD enriched control sample by demanding these values to be small.
The ATLAS collaboration has also presented jet smearing techniques to estimate the remaining QCD background as 
can be found in Ref.~\cite{atlas}. Jet transverse momenta in low $E_{T,\rm{miss}}$ QCD multijet data is 
smeared with a jet response function $R$ measured from data. The Gaussian part of $R$ is measured from
$\gamma$+jet events and
the non-gaussian part 
from mercedes type 3-jet events where the missing transverse momentum can be
unambiguosly associated with one jet. Determination of the QCD background will be
a challenge and only a comparison of a few uncorrelated techniques will give a robust estimate.
\subparagraph*{Background from top and $W$ events}
Top and $W$ background can be estimated in a sample where one lepton is required in addition to the jet and $E_{T,\rm{miss}}$ 
cut. Top and $W$ events are predominantly found in such events if the transverse mass
between the lepton and the missing momentum vector is less than $M_W$ as can be seen in Figure~\ref{mt}. 
Due to resolution effects a cut 
$M_T<100$~GeV was required to define the control region. 
The $M_T>100$~GeV selection is the one lepton signal region.
Although the composition of the different type of top decay modes and the fraction of $W$ events differ between
signal and control regions, both types of events have similar kinematical distributions. Therefore the control sample is able to
predict the $M_{\rm{eff}}$ distribution in the signal region within statistical precision as can be seen in 
Figure~\ref{mt}. If the lepton composition and efficiency is known, the same method can determine the $M_{\rm{eff}}$ 
distribution in the 0-lepton channels.
In these methods, the $M_T$ cut efficiency also needs to be estimated. 
\subparagraph*{Background from $Z$ events}
A straightforward method to determine the background from $Z \rightarrow \nu \overline{\nu} $+jets events is to measure 
$Z \rightarrow e \overline{e} $ and $Z \rightarrow \mu \overline{\mu} $+jets from data and to correct these measurements
for the lepton identification efficiencies, the branching ratios and the unmeasured phase space at low $p_T$ and high pseudorapidity of the leptons. This method works effectively and provides a nice estimate up to medium $M_{\rm{eff}}$ as can 
be seen in Figure~\ref{z}. 
At high $M_{\rm{eff}}$ the statistics of the lepton control samples is too low due to the smaller branching ratio of the 
$Z$ to leptons. 
The CMS experiment has studied the determination of this background from $W$+jets and $\gamma$+jets events since
similar shapes for all types of bosons are expected at high boson transverse momentum \cite{cmsz}. 
The photon control selection is very promising and provides high statistics with a good signal to background ratio. 
The good agreement of the estimated and real $Z$ background is shown in Figure~\ref{z}. 
An alternative measurement can be made from W+jet events. 
Here the high statistics is counterbalanced by large backgrounds
from top events and a possible signal contamination.
\subparagraph*{Reach in SUSY parameter space}
The insight gained from these studies determine uncertainties for each background
source. These uncertainties have been applied to define ATLAS and CMS significance in the SUSY parameter
space. The significance is calculated for each SUSY point 
without changing the selections from SUSY point to SUSY point to ensure generality and statistical coherence. 
ATLAS used a set of 5 cuts on $M_{\rm{eff}}$ to choose the best significance and corrected this significance 
for the effect of 5 statistical tests. CMS used its default selections without re-optimisation. 
Figure \ref{reach} shows the 5$\sigma$ discovery reach for the mSUGRA model with $\tan \beta =10$,
$A_0=0$~GeV and $\mu=+$ for various channels at the ATLAS and CMS experiments. It is remarkable that both experiments
find a very similar reach in parameter space. ATLAS compared different jet multiplicities and found in some regions
a better discovery potential for 2- and 3-jet selections.
ATLAS determined the reach in parameter space as well for mSUGRA constrained from measurements including the
dark matter density, for a model with non-universal higgs masses (NUHM), for mSUGRA with high $\tan \beta$ and gauge mediated SUSY breaking (GMSB). All scans have been performed with a fast parameterization of the detector response.
The results of these scans and the studies with the full simulation indicate that both ATLAS and CMS should discover
signals of R-parity conserving SUSY with gluino and squark masses up to 1 TeV after having accumulated and understood
$1\rm{fb}^{-1}$ of data. 

\section{EXCLUSIVE MEASUREMENTS}
\begin{figure*}[t]
\centering
\includegraphics[width=47mm]{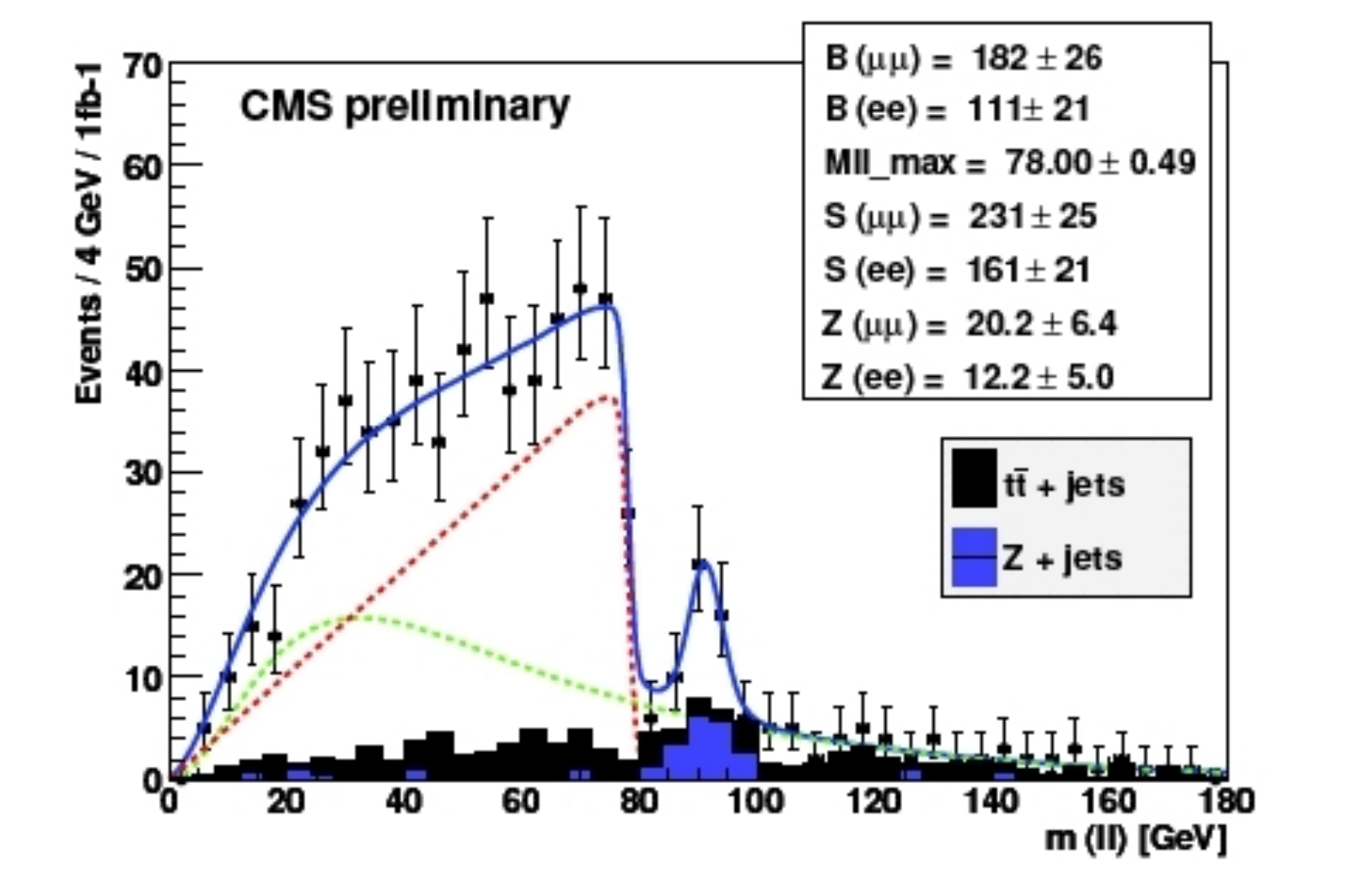}
\caption{The invariant mass distribution of the CMS dilepton selection.} \label{exclusive}
\end{figure*}

After a signal with missing transverse momentum has been observed we
would need to determine if SUSY is the source of the excess events.
To answer this question requires many exclusive measurements. 
One important example is the measurement of the $\chi_2^0 \rightarrow \tilde{l}_{L,R} l \rightarrow l l \tilde{\chi_1^0}$
decay in opposite sign dilepton events \cite{cmsexclusive}. Due to missing energy no mass peaks are observable, but the shapes and endpoints
of the dilepton mass distribution provide mass information. 
CMS determined the same flavour top and di-boson background from $e\mu$ data. 
The dilepton endpoint was then found by a 6 parameter fit to the $ee$ and $\mu \mu$ mass distributions. The result of such
a fit can be seen in Figure \ref{exclusive}. The statistical accuracy of the endpoint of LM1 was found within a total
uncertainty of 1 GeV. A similar analysis is performed in ATLAS \cite{atlas}. 
\begin{acknowledgments}
The author acknowledges the support by the Landesstiftung Baden W\"urttemberg and the BMBF.
\end{acknowledgments}

\end{document}